\magnification 1200

\rightline{KCL-MTH-10-16}

\vskip .5cm
\centerline {\bfÊ  $E_{11}$, generalised space-time and IIA string
theory; the $R\otimes R$ sector }

\vskip 1cm
\centerline{Andreas Roc\'{e}n and Peter West}
\centerline{Department of Mathematics}
\centerline{King's College, London WC2R 2LS, UK}

\vskip .5 cm
\noindent
We extend the analysis of arXiv:1009.2624,  which constructed the
non-linear realisation of the semi-direct product of $E_{11}$ and the
$l_1$ representation at level zero,  to level one. Thus we add to the
previously considered $NS\otimes NS$ fields those of the $R\otimes R$
sector. 
\vfillÊ
\eject

\medskip
Since the original suggestion of an $E_{11}$ symmetry of the underlying
theory of strings and branes [1] there has accumulated a
substantial body of evidence in its support.  However, it
still remains to incorporate space-time in a way that is consistent with
$E_{11}$ and  leads to the theories with which we are familiar. It has
been suggested that one considers the non-linear realisation of the
semi-direct product of
$E_{11}$  and  its  first fundamental representation, denoted
$E_{11}\otimes_s l_1$ [2]. This introduces a generalised space-time which
contains coordinates for each field, for example in eleven
dimensions in addition to the  usual coordinates $x^a$ associated with the
metric we also have the coordinates $x^{ab}$ associated with the three
form field as well as coordinates associated with all the higher
level fields that the non-linear realisation of $E_{11}$ contains.  The
non-linear realisation determines the dynamics of the fields, up
to some constants,  even though they depend on the additional
coordinates and so it also specifies the  extension of
Einstein's geometry that must occur. However, the theories we are used
to  contain fields which just depend on the usual coordinates
$x^a$ and what is not clear is how they can be recovered from the
formulation involving the generalised space-time. Nonetheless, the
non-linear realisation of 
$E_{11}\otimes_s l_1$ has been used to derive many of the properties of
gauged supergravities [3]. This latter reference also includes a summary of
the evidence for $E_{11}$ and that for the $l_1$ representation as the
multiplet of all brane charges. 

Recently   [4] the level zero non-linear
realisation of $E_{11}\otimes_s l_1$,  from the IIA perspective,  was
constructed. It lead to an $O(10,10) \otimes GL(1)$ invariant field
theory living on a space-time that transformed as the 20 dimensional
vector of O(10,10). The field content in this generalised space-time was
that of the $NS\otimes NS$ sector of the superstring.  We refer the
reader to this paper for the details of this calculation,  the
conventions employed and references to the literature. The result agreed
 with the doubled field theory of reference [5] which was derived from
closed field theory and other considerations. In this short note
we extend the calculation of reference [4] to level one whose fields are
contained in a Majorana-Weyl spinor of O(10,10). Indeed it  contains all
the totally anti-symmetric fields of odd rank and they are those  of the
$R\otimes R$ sector.
\par
The non-linear realisation at level zero is for the group $O(10,10)
\otimes GL(1)$ with local subalgebra $O(10)\times O(10)$. The local
subalgebra will play an important role in our construction and in order
to find its action it will be useful
to reformulate the discussion of [4]  in  a manifestly  O(10,10)
formulation. Let us denote the level zero generators of the
$l_1$ representation by 
$P_A= (P_a, \bar P^{\bar a}), A=1,\ldots ,20$,  we see that the index
$A$ is subdivided  $A= (a,\bar a), a=1,\ldots ,10, \
\bar a=1,\ldots ,10$. Alternatively, we may write this equation in the
form  
$$
P_A=\cases{ P_a\quad a=A,\  a= 1,\ldots , 10\cr 
\bar P^{\bar a} \quad \bar a= A-10,\ \bar a = 1,\ldots , 10}
\eqno(1)$$
In terms of the  notation of reference [4], $P_a=P_a$ but $Q^a= 2\bar
P^{\bar a}$. The corresponding coordinates are denoted by $Z^A=(z^a, \bar
z_{\bar a} )$. In terms of our previous coordinates $z^a=x^a, \bar
z_{\bar a}=2 y_a$. The generators $P_A$ act on the coordinates by
$P_A={\partial\over
\partial Z^A}$. The line element 
$d^2s= 2dz^ad\bar z_{\bar a}\equiv  dZ^A\Omega _{AB}dZ^B$ 
  is O(10,10) invariant. The object $\Omega _{AB}= \Omega ^{AB}$ is
the O(10,10) invariant metric and  can be used to raise and
lower indices, for example
$P^A=\Omega^{AB}P_B=(P^{\bar a}, P_a)$. We note that the unbarred
indices, $a,b,\ldots$,   always appear in the same position to the
SO(10,10) indices  $A,B,\ldots $, but the barred indices,
$\bar a,\bar b,\ldots$,    appear in the opposite position. Also 
the metric takes the values  $\Omega^a{}_{\bar
b}=\delta^a_b$, $\Omega_{\bar a}{}^b=\delta_a^b$,
$\Omega^{ab}=0=\Omega_{\bar a\bar b}$
\par
We denote the generators of O(10,10)  by
$L^{A B}$ and their  action on the coordinates is given by the operator  
$ L^{A B}= Z^A\Omega^{BC}{\partial\over \partial
Z^C}-Z^B\Omega^{AC}{\partial\over \partial Z^C}$. They obey the commutator
$$
[ L^{AB}, L^{CD}]= \Omega ^{BC} L^{AD}+\Omega ^{AD} L^{BC}-\Omega ^{AC}
L^{BD}-\Omega ^{BD} L^{AC}
\eqno(2)$$
In terms of our previous notation for the O(10,10)  generators $L^A{}_B$
are given by 
$$
L^a{}_b=\tilde K^a{}_b,\ L^{a}{}^{\bar b}= R^{ab},\ L_{\bar
a}{}_b=-R_{ab},\  L_{\bar a}{}^{\bar b}=-\tilde K^b{}_a
\eqno(3)$$
and we recover equation (A.11).  
\par
The commutation relations of the O(10,10) generators with $P_A$
are given by 
$$
[ L^{AB},P_C]= -\delta_C^A \Omega^{BD}P_D+\delta_C^B \Omega^{AD}P_D
\eqno(4)$$
which is equivalent to equation (A.18). 
\par
We now introduce the matrices 
$\beta^A=( \beta^a, \bar \beta_{\bar a})$ which obey
the relations 
$$
\{\beta^A, \beta^B\}=\Omega^{AB}
\eqno(5)$$
The generators of O(10,10)  act on spinors by taking 
$$
L^{AB}= {1\over 2} [\beta^A,\beta^B]= \beta^A \beta^B-{1\over 2}
\Omega^{AB}
\eqno(6)$$
We can write a spinor as a polynomial of  $ \beta^a$ acting on a vacuum
$|0>$ which obeys $\bar \beta_{\bar a} |0>=0$; 
$$
|\varphi >= (\phi+  A_a \beta^a +\ldots +{A_{a_1\ldots a_r}\over r!}
\beta^{a_1}\ldots \beta^{a_r} 
  +\ldots )|0>
\eqno(7)$$
\par
We now make the change of basis from $P_A$ to $Q_A= (Q_a, \bar Q_{\bar
a})$ where 
$$
Q_a=P_a+\bar P^{\bar a},\  \bar Q_{\bar a}=-P_a+\bar P^{\bar a}\quad
{\rm or} \quad P_a= {1\over 2}(Q_a-\bar Q_{\bar a}),\ \bar P^{\bar a} =
{1\over 2}(Q_a+\bar Q_{\bar a})
\eqno(8)$$
This change corresponds to introducing the coordinate change from $Z^A$
to  
$Y^A= ( y^a,
\bar y^{\bar a}) $,  not to be confused with the coordinate $y_a$ used
in reference [4],  and they are related by 
$$
y^a={1\over 2}( z^a+\bar z_{\bar a}), \  \bar y^{\bar a}={1\over
2}(-z^a+\bar z_{\bar a})\quad {\rm or} \quad z^a= (y^a-\bar
y^{\bar a}), \ \bar z_{\bar a} = (y^a+\bar y^{\bar a})
\eqno(9)$$
The invariant line element is given by $d^2 s= 2dz^ad\bar z_{\bar
a}= 2(dy^ady^a-d\bar y^{\bar a}d\bar y^{\bar a})=G_{AB}dY^AdY^B$ where the
invariant metric
$G_{AB}={\rm diag} (2,\ldots ,2, -2,\ldots ,-2)=4 G^{AB}$ is now diagonal
but has signature zero. In this new basis both barred and unbarred
indices have the same position as the SO(10,10) indices. 
We note that
$Q_A={\partial\over
\partial Y^A}$. 
\par
The $\gamma$ matrices  in
the new basis, $\gamma^A= (\gamma^a, \bar \gamma^ {\bar a} )$ are given by 
$$
\gamma^a= ( \beta^a+\bar \beta_{\bar a}), \  \bar \gamma^{\bar
a}= (-\beta^a+\bar \beta_{\bar a})\quad {\rm or} \quad 
\beta^a= {1\over 2}(\gamma^a-\bar
\gamma^{\bar a}), \ \bar \beta_{\bar a} = {1\over 2}(\gamma^a+\bar
\gamma^{\bar a})
\eqno(10)$$
We have scaled the $\gamma^A$ by a factor of two from that which the
straightforward change of basis would imply. 
They obey 
$$
\{\gamma^A, \gamma^B\} =  4 G^{AB}
\eqno(11)$$
which we recognise as the usual Clifford algebra
relations in signature $(10,10) $ space-time. For convenience we have
scaled the $\gamma^A$ by a factor of two. 
\par
We denote the O(10,10) generators in the
new basis by $M^{AB}$ and in spinor basis they are given, including
their normalisation, by
$M^{AB}= {1\over 4} [\gamma^A,\gamma^B]$.
The components of $M^{AB}$ are most straightforwardly evaluated by
using its spinor representation. We find that 
$$
M^{ab}=  {1\over 4} [\beta^a, \beta ^b]+{1\over 4} [\beta^a,
\bar \beta _{\bar b}]+ {1\over 4} [\bar \beta_{\bar a}, \beta ^b]+ {1\over
4} [\bar \beta_{\bar a},\bar  \beta _{\bar b}]={1\over
2} (J^{ab}+R^{ab}- R_{ab})
\eqno(12)$$
where $J^{ab}= \tilde K^a{}_c\eta^{cb}-\tilde K^b{}_c\eta^{ca}$ which  we
recognise as the generators of the  Lorentz group.  Similarly we find
that 
$$
M^{\bar a\bar b}=- {1\over 2} (J^{ab}-(R^{ab}-
R_{ab}))
\eqno(13)$$
\par
The advantage of the change of basis now become clear. It is
straightforward to show that
$[M^{ab}, M^{\bar c\bar d}]=0$ and indeed 
$ M^{ab}$ and $ M^{\bar a\bar b}$ generate the algebra $O(10)\otimes
O(10)$. This is none other than the Cartan involution invariant algebra of
O(10,10) which, as we will see,  plays a crucial role in the construction
of the dynamics. In this basis $Q^A$ splits into the $(10,0)+ (0,10)$
representations  of
$O(10)\otimes O(10)$; indeed 
$$
[M_{ab}, Q_c]= \delta _{bc} Q_a -\delta _{ac} Q_b ,\ [M_{\bar a \bar b},
 Q_{ c}]=0,\ [M_{ab}, \bar Q_{\bar c}]=0,
\ [- M_{\bar a \bar b},
 \bar Q_{\bar  c}]=\delta _{\bar b\bar c} \bar Q_{\bar a} -\delta _{\bar
a\bar c} \bar Q_{\bar b}
\eqno(14)$$
\par
The analogue of $\gamma_5$ for O(10,10) is given by 
$$
\Gamma\equiv  -\gamma^1\gamma^2\ldots \gamma^{10} \bar
\gamma^{\bar 1}\bar \gamma^{\bar 2}\ldots
\bar \gamma^{\bar {10}}= \gamma^1\bar \gamma^{\bar 1}\gamma^2\bar
\gamma^{\bar 2}\ldots \gamma^{10}\bar \gamma^{\bar {10}}
\eqno(15)$$
As we are  in a space with signature zero we can impose spinors to be
simultaneously Majorana and Weyl. Indeed we demand that 
$$\Gamma |\varphi >= -|\varphi>
\eqno(16)$$ 
implying  that the spinor $\varphi$ contains only odd forms. Thus it does
contain the fields of the $R\otimes R$ sector. 
\par
Finally we can construct the dynamics,  that is the non-linear realisation
of $E_{11}\otimes l_1$ at level one. We may write the group element 
as 
$$
g= e^{z^aP_a+\bar z _{\bar a} \bar P^{\bar a}}e^{ K^a{}_b h_a{}^b}
  e^{ {A_{a_1a_2}\over 2} R^{a_1a_2}} e^{<\bar R |\varphi >}
 e^{aR}
\eqno(17)$$
The third  factor can be written as 
$$
e^{<\bar R |\varphi >}= e^{ R^a A_a+\ldots
+R^{a_1\ldots a_r} {{A_{a_1\ldots a_r}\over r!}}+\ldots } \eqno(18)$$
provided we identify  
$$
< \bar R |= <0| (\bar \beta_{\bar a} R^a+\ldots +(-1)^{{r(r-1)\over 
2}}\bar
\beta _{\bar a_1}\ldots\bar \beta_{ \bar a_r} {R^{a_1\ldots a_r}\over
r!}+\ldots) 
\eqno(19)$$
where $ <0|\beta^a=0$. We note that we could have written  
$<\bar R|\varphi>=\bar R^\alpha
\varphi_\alpha$; that is,  as a more usual spinor contraction. Since we
only want generators of odd rank we impose the Weyl condition   
$$
 <\bar R|\Gamma= -<\bar  R|
\eqno(20)$$
Up to field redefinitions and rescaling of generators by constants the
above group element is the same as that used to compute  the field
strengths previously  in papers on the IIA theory from the $E_{11}$
perspective, except that we now have the
dependence on the extra coordinate
$\bar z_{\bar a}$. We have used the GL(10) generators $K^a{}_b$
of the $E_{11}$ algebra rather than the $\tilde K^a{}_b$  that occur in
the SO(10,10) subalgebra; the relation between the two being 
$\tilde K^a{}_b= K^a{}_b +{1\over 6} \delta_b^a \tilde R$ with 
$\tilde R= 3R -{3\over 4} \sum_{a=1}^{10} K^a{}_a$. 
\par
The Cartan forms are given by 
$$
{\cal V} = g^{-1} d g= dZ^N E_N{}^A P_A+ dZ^N{\cal V}^{(0)}_N + dZ^N\bar
R^\alpha D_N\varphi_\alpha= dZ^N E_N{}^A (P_A+ {\cal V}^{(0)}_A + \bar
R^\alpha D_A\varphi_\alpha )
\eqno(21)$$
where $ E_N{}^A $ is the generalised vielbein  given by 
$$
 E_N{}^A = (\det e )^{-{1\over 2}}\left(\matrix {e&
eAe^{-{1\over 2}a}\cr 0&e^{-1T}e^{-{1\over 2}a}\cr}\right)
\eqno(22)$$
$e_\mu{}^a= (e^h)_\mu{}^a$ and  $\partial_N={\partial\over \partial Z^N}$.
We have now adopted the convention that tangent indices are denote by
$A,B,\ldots$ while world indices are denoted  by
$M,N,\ldots $. The  object $ {\cal V}^{(0)}_A$ is the Cartan form of 
O(10,10) and it is given by 
$$
{\cal V}^{(0)}_A= E_A{}^N((e^{-1}\partial_N e)_a{}^b K^a{}_b
+e^{-{a\over 2}}{1\over 2} \tilde D_N A_{a_1a_2}R^{a_1a_2}+\partial_N a)
\eqno(23)$$
where $\tilde D_N A_{a_1a_2} = \partial_N A_{a_1a_2} +(e^{-1}\partial_N
e)_{a_1}{}^b A_{b a_2 }+(e^{-1}\partial_N e)_{a_2}{}^b
A_{a_1 b}$ with similar covariantizations when acting on other fields. 
The Cartan form for the $R\otimes R$ fields is given by 
$$
 D_A |\varphi> = e^{-a({1\over 4}
\sum_a \beta^a \bar \beta_{\bar a}-1)}  E_A{}^N(\tilde D_N \varphi+
 {1\over 2}\tilde D_N A_{a b} \beta^a\beta^b )|\varphi>
\eqno(24)$$
In deriving this result we have used that 
$$
[K^a{}_b, <\bar R | ]=  <\bar R | \beta ^a
\bar \beta_{\bar b},\quad
[R^{ab}, <\bar R | ]=  <\bar R | \beta ^a \beta^ b
,\quad [R,<\bar R | ]= <\bar R |
({1\over 4}\sum_a \beta^a \bar \beta_{\bar a}-1)
\eqno(25)$$
A comparison of the commutation relation derived from this equation with
those of  reference  [2] implies that the generators here and those in
that paper, now denoted with a tilde,  are related by 
$R^{a_1a_2}=\tilde R^{a_1a_2}$ and
$R^{ a_1\ldots a_p} = e_p\tilde R^{ a_1\ldots a_p}
$ with $e_1=1, e_3=-2, e_5=-4, e_7=-8, e_9=4.8$. 
\par 
 The Cartan form is
invariant under the rigid O(10,10) transformations of the non-linear
realisation but transforms under  local $h\in O(10)\otimes O(10)$
transformations as
${\cal V}^\prime= h^{-1}{\cal V}h+h^{-1}d h$. At the level at which we are
working the  $D_A\varphi_\alpha $ part of the Cartan form transforms
covariantly.  As we found above the $O(10)\otimes O(10)$ transformations
are most easily written down in the $Y^N$ coordinate basis. Let us denote
the derivative contracted with  the generalised vielbein in this basis as 
$$
\Delta _A=(\Delta_a,\bar \Delta_{\bar a})=( D_a+\bar
D^{\bar a}, -D_a+\bar D^{\bar a})
\eqno(26)$$ 
Then $\Delta_a$ transforms 
as a vector under  the  first of the O(10) factors  and is inert under
the second and vice-versa  for $\Delta_{\bar a} $. As such there   are 
two possible covariant terms  which are  first order in derivatives; 
$$
c_1 \gamma^a\Delta _a |\varphi >+c_2  \bar \gamma^{\bar a}\bar \Delta
_{\bar a }|\varphi >
\eqno(27)$$
\par
As we did in reference [4] at level zero, we can demand that if we set
${\partial\bullet
\over \partial\bar z_{\bar n}}$ where $\bullet$ is any field then the
result should be gauge invariant. The operator  acting on $\varphi$ in
the above equation can be rewritten so that it has two types of
terms, one of  which contains $\bar
\beta_{\bar a}$ and the other 
$\beta^a$  both of which multiply derivatives. At the linearised
level these have the effect of giving the divergence and curl of the
fields respectively. Clearly, we would like the curl so as to be gauge
invariant. This requires
$c_2=c_1={1\over 2}$ where in the last equality we have chosen the
scale. Thus the object 
$$
E\equiv {1\over 2}  \gamma^a\Delta _a |\varphi>+{1\over 2}   \bar
\gamma^{\bar a}\bar\Delta _{\bar a }|\varphi>= (\beta^a
D_a+\bar\beta_{\bar a} \bar D^{\bar a})|\varphi>
\eqno(28)$$
is inert under rigid O(10,10) transformations,  transforms like a
spinor under $O(10)\otimes O(10)$ transformations.  
\par
The object 
$$
\gamma=\gamma^1\gamma^2\ldots \gamma^{10}
\eqno(29)$$
is the Weyl operator associated with  the first of the O(10) factors
in  $O(10)\otimes O(10)$. As such an invariant equation of motion  is
given by 
$$
\gamma E=E
\eqno(30)$$
If we set the  derivatives with respect to $\bar z^{\bar n}$
to zero then $E$ is  just contains the decorated field strengths  of the
IIA theory,  as they usually appear in the calculation of the non-linear
realisations of
$E_{11}$. The above 
 equation just sets the field strengths equal to their duals
using the space-time epsilon symbol. These are   the correct
equations of motion  for the fields in the $R\otimes R$ sector; that is,
they are those that occur in the IIA supergravity theory. 
\par
Indeed, using equation (24) we find that if the fields are independent of
$\bar z_{\bar a}$ and we define $E=\beta^aD_a |\varphi> = \sum_r
\beta^{a_1}\ldots \beta ^{a_r}{F_{a_1\ldots a_r}\over (r-1)!}|0>$ then 
$$
F_{a_1a_2}= e^{{3a\over 4}}2(\partial_{[a_1}A_{a_2]}), \ 
F_{a_1\ldots a_4}= e^{{a\over
4}}4(\partial_{[a_1}A_{a_2a_3a_4]}+3\partial_{[a_1}A_{a_2a_3}A_{a_4]}), \ 
$$
$$
F_{a_1\ldots a_6}= e^{-{a\over
4}}6(\partial_{[a_1}A_{a_2\dots
a_6]}+2.5\partial_{[a_1}A_{a_2a_3}A_{a_4\ldots a_6]}),
$$
$$
F_{a_1\ldots a_8}= e^{-{3a\over
4}}8(\partial_{[a_1}A_{a_2\dots
a_8]}+3.7\partial_{[a_1}A_{a_2a_3}A_{a_4\ldots a_8]}),
$$
$$
F_{a_1\ldots a_{10}}= e^{-{5a\over
4}}10(\partial_{[a_1}A_{a_2\dots
a_{10}]}+4.9\partial_{[a_1}A_{a_2a_3}A_{a_4\ldots a_{10}]})
\eqno(31)$$
These look rather different to those previously derived, say in reference
[6]. Corresponding to the rescaling of the generators below equation (25) 
we must scale the   fields  by $ (e_p)^{-1} $, but  we must also 
take into account  the difference in the ordering of the factors in the
group element of equation (17) from that of  reference [6]. The
effect is that the   fields 
here and those in reference [6], which are denoted with a tilde  are
related by 
$$
A_{a_1a_2}= \tilde A_{a_1a_2},\ A_{a_1}= \tilde A_{a_1},\
A_{a_1a_2a_3}=-{1\over 2} \tilde A_{a_1a_2a_3},\  A_{a_1\ldots a_5}
= {1\over 4}(\tilde A_{a_1\ldots a_5}+20\tilde A_{a_1a_2}
\tilde A_{a_3\ldots a_5})$$ 
$$
A_{a_1\ldots a_7} = {1\over 8}(\tilde A_{a_1\ldots a_7}-76
\tilde A_{a_1a_2}\tilde A_{a_3\ldots a_7}-7.6.10
\tilde A_{a_1a_2} \tilde A_{a_3a_4} \tilde A_{a_5\ldots a_7})
$$ 
$$
A_{a_1\ldots a_9} = -{1\over 8.4}(\tilde A_{a_1\ldots a_9}-144(
\tilde A_{a_1a_2}\tilde A_{a_3\ldots a_9}-7.3
\tilde A_{a_1a_2} \tilde A_{a_3a_4}\tilde A_{a_5\ldots a_9} -7.3.10
\tilde A_{a_1a_2}\tilde A_{a_3a_4}\tilde A_{a_5a_6}\tilde A_{a_7\ldots
a_9}))
\eqno(32)$$
Substituting these into the field strengths of even rank of equation (31)
we find precisely those of equation (1.24-1.32) of reference [6]. 
\par
We could have included the coordinates at level one which
belong to a  Majorana-Weyl spinor of the opposite chirality to $\varphi$.
These will lead to a larger generalised vielbein which has four blocks;
the upper diagonal block is the generalised vielbein which we have given above
in equation (22), the lower diagonal block is constructed from  vielbeins,
the lower left block is zero and the upper right block has the generic
form $E_N{}^A (\gamma_A\varphi)_\alpha$. The inverse generalised vielbein
also has the same generic form.  The derivatives in the generalised
space-time will occur in the dynamics contracted with the world index of 
the larger inverse generalised vielbein in order to have an object that
transforms under local transformations, that is generalised tangent 
space rotations. As such the derivatives above will gain an additional
term that is proportional to
$\varphi$ times the derivative with respect to the new coordinates,
however, this term is of level two and so of higher level than needed in
this paper.

\medskip 
{\bf References}
\medskip 
\item{[1]} P. West, {\it $E_{11}$ and M Theory}, Class. Quant.
Grav. {\bf 18 } (2001) 4443, {\tt hep-th/0104081}
\item{[2]} P. West, {\it $E_{11}$, SL(32) and Central Charges},
Phys. Lett. {\bf B 575} (2003) 333-342, {\tt hep-th/0307098}
\item{[3]} F. Riccioni and P. West, {\it E(11)-extended space-time and
gauged supergravities}, JHEP0802:039,2008;  hep-th/0712.1795
\item{[4]} P. West, $E_{11}$, Generalised space-time and IIA string
theory, arXiv:1009.2624.
\item{[5]} O. Hohm, C. Hull and B. Zwiebach, {\it Generalised metric
formulation of double field theory},  hep-th/1006.4823. 
\item{[6]}  I. Schnakenburg and P. West, {\it Massive IIA Supergravity as
a Non-linear Realisation},   Phys.Lett. B540 (2002) 137-145,
hep-th/0204207.

\end